\documentclass[aps,pra,twocolumn,showpacs,groupedaddress,amsmath,amssymb]{revtex4}
\usepackage{graphicx}
\begin{document}
\title{Magnetism in a lattice of spinor Bose condensates}

\author{Kevin Gross, Chris P. Search, Han Pu, Weiping Zhang and Pierre
Meystre}
\affiliation{ Optical Sciences Center, The University of
Arizona, Tucson, AZ 85721}
\date{\today}

\begin{abstract}
We study the ground state magnetic properties of ferromagnetic
spinor Bose-Einstein condensates confined in a deep optical
lattices. In the Mott insulator regime, the ``mini-condensates''
at each lattice site behave as mesoscopic spin magnets that can
interact with neighboring sites through both the static magnetic
dipolar interaction and the light-induced dipolar interaction. We
show that such an array of spin magnets can undergo a
ferromagnetic or anti-ferromagnetic phase transition under the
magnetic dipolar interaction depending on the dimension of the
confining optical lattice. The ground-state spin configurations
and related magnetic properties are investigated in detail.
\end{abstract}
\maketitle

\section{Introduction}

The interaction between quantum degenerate atomic gases and
optical fields is a cornerstone of modern atomic physics and
quantum optics. In early experiments on Bose-Einstein
condensation, light fields were applied primarily for the capture
and precooling of atoms, preceding the last stage of evaporative
cooling in a magnetic trap. They were also used to obtain dramatic
images of the condensates (BEC), and to launch solitons
\cite{solitons} and vortices \cite{vortices} in condensates. This
was soon followed by applications such as the trapping of
condensates in optical dipole traps, and the demonstrations of
matter-wave superradiance \cite{superad,moore} and of coherent
matter-wave amplification \cite{amplify}. More recently, optical
dipole traps have been employed for the all-optical realization of
BEC and of quantum-degenerate Fermi gases \cite{chap,thomas}.

Optical lattices, formed by counterpropagating laser beams in one,
two and three dimensions, were originally used in polarization
gradient cooling and sub-recoil cooling experiments at the
single-atom level. They rapidly found further applications in the
manipulation of BECs, first in the demonstration of a
``mode-locked" atom laser and the observation of Josephson
tunneling between lattice wells \cite{yale1}, and subsequently in
the transport and acceleration of condensates \cite{accelerate}.
More recently, they have lead to the demonstration of the
superfluid-Mott insulator transition \cite{mott,zoller}, and of
the collapse and revival of the condensate wave function
\cite{collapse}. In the near future, they may also prove useful in
the realization of bright atomic solitons relying on negative
effective atomic masses in periodic potentials \cite{bright}.

In contrast to magnetic traps, which only capture atoms in
weak-field seeking states, optical traps function for all
hyperfine sublevels of the alkali electronic ground states. This
presents considerable advantages, in particular in the study of
spinor condensates such as sodium and rubidium. The first study of
the magnetic properties of spinor condensates were carried out by
Ketterle and coworkers, who investigated the existence of
coexisting spin domains in $^{23}$Na, an ``anti-ferromagnetic'',
or ``polar'' condensate \cite{mit}.

Recent experimental and theoretical studies have established that
in contrast to $^{23}$Na, $^{87}$Rb is expected to be
ferromagnetic at zero temperature. That is, the expectation value
of its total spin ${\bf F}$ is finite, $\langle {\bf F} \rangle
\neq 0$ \cite{green,klausen,heinzen}. As a result, an ensemble of
condensates placed at the potential minima of an optical lattice
would act as mesoscopic magnets, much like large spins on a
crystalline lattice. In the absence of external fields and long
range site-to-site interactions, these magnets would have random
orientations.

The situation is changed in the presence of interactions between
neighboring lattice sites. It is known that in the case of spins
on a crystal lattice, the dominant source of coupling is the
quantum-mechanical exchange interaction. We recall that 19th
century physics failed in its attempts to explain ferromagnetism
in terms of the magnetic dipole-dipole interaction, and it is
Heisenberg who first introduced the exchange force to explain this
effect \cite{huang,solid}. In the present case, though, the
overlap between neighboring condensate wave functions is
negligible for deep enough lattice wells --- the Mott insulator
state --- and so is the exchange interaction. Instead, the
individual mesoscopic magnets are coupled by the magnetic (and
possibly also the optical) dipole-dipole interaction. Because of
the large number $N$ of atoms at each lattice site, this
interaction is no longer negligible, despite the large distance,
of the order of half an optical wavelength, between sites. As
such, the present situation is in some sense a return to 19th
century physics. The goal of this paper is to discuss several
aspects of the spin and magnetic properties of such lattice
systems in one- and two-dimensions.

The remainder of the paper is organized as follows. Section II
briefly reviews the theory of spinor condensates in general, with
special emphasis on the ferromagnetic and polar ground states
resulting from local spin-changing collisions. We then introduce
the nonlocal long-range magnetic dipole-dipole interaction between
condensates at different sites in the optical lattice. Section III
briefly reviews previously published results on one-dimensional
lattices, and discusses the ferromagnetic ground state of the full
lattice. On this basis, an extension of the one-dimensional case
to two-dimensional lattices are analyzed in section IV. The ground
state of the system is determined numerically using a genetic
algorithm that is discussed in some detail. We show that in that
case, the ground state is normally anti-ferromagnetic. Edge
effects are also briefly addressed. Finally, Section V is a
conclusion and outlook.

\section{model}

The dynamics of spinor condensates trapped in optical lattices is
primarily governed by three types of two-body interactions:
spin-changing collisions, magnetic dipole-dipole interactions, and
light-induced dipole-dipole interactions. For an optical lattice
created by blue-detuned laser beams, the atoms are trapped in the
dark-field nodes of the lattice and the light-induced
dipole-dipole interaction can be neglected \cite{note}. In this
paper, we focus on this case. As a preparation for sections III,
IV and V, we first discuss the interatomic interactions in some
detail.

\subsection{Spin-changing collisions}

In second-quantized form, the Hamiltonian describing a system of
spin $f=1$ bosons subject to local spin-changing collisions is
\cite{zhang,ho2,ohmi,law}
\begin{eqnarray}
{\cal H}&=&\sum_\alpha \int d^3r \left ( \frac{\hbar^2}{2M}\nabla
\psi_\alpha^\dagger({\bf r}) \cdot \nabla \psi_\alpha({\bf r}) +
U({\bf r})\psi_\alpha^\dagger \psi_\alpha({\bf r}) \right )\nonumber \\
&+& \frac{c_0}{2}\sum_{\alpha,\beta} \int d^3r \psi_\alpha^\dagger
({\bf r})\psi_\beta^\dagger ({\bf r})\psi_\beta({\bf
r})\psi_\alpha({\bf r})  \\&+&
\frac{c_2}{2}\sum_{\alpha,\beta,\mu,\nu} \int d^3 r
\psi_\alpha^\dagger ({\bf r})
 \psi_\beta^\dagger({\bf r}) {\bf
F}_{\alpha,\mu} \cdot {\bf F}_{\beta,\nu }\psi_\nu({\bf
r})\psi_\mu({\bf r}),\nonumber  \label{hamiltonian}
\end{eqnarray}
where $\psi_\alpha({\bf r})$ is the field annihilation operator
for an atom in the hyperfine state $|f=1,m_f=\alpha\rangle$,
$\alpha = -1,0,1$, $U({\bf r})$ is a potential produced by an
optical dipole trap and assumed to be the same for all hyperfine
states, and $M$ is the mass of the atoms. ${\bf F}$ is the vector
operator for the hyperfine spin of an atom, with components
represented by $3\times 3$ matrices in the
$|f=1,m_f=\alpha\rangle$ subspace. For ultracold bosons, only
$s$-wave collisions with total hyperfine spin of $F=0,2$ are
allowed, and
\begin{equation}
c_0= \frac{4\pi \hbar^2}{3M} \left(a_0 + 2 a_2 \right ), \nonumber
\end{equation}
and
\begin{equation}
c_2=\frac{4\pi \hbar^2}{3M} \left(a_2 - a_0 \right ), \nonumber
\end{equation}
where $a_0$ and $a_2$ are the $s$-wave scattering lengths for
collisions in the $F=0$ and $F=2$ channel, respectively.

The ground state properties of spinor condensates subject to these
local spin-changing collisions have been determined by introducing
the components $\phi_a({\bf r})$ of the spinor condensate wave
function in the mean-field approximation,
\begin{equation}
\phi_\alpha({\bf r}) = \langle \psi_\alpha ({\bf r}) \rangle =
\sqrt{n({\bf r})} \zeta_\alpha({\bf r}),
\label{wf}
\end{equation}
where $n({\bf r})$ is the local atomic density and
$\zeta_\alpha({\bf r})$ a normalized spinor, and minimizing the
energy functional
\begin{eqnarray*}
E &=& \int d^3r \frac{\hbar^2}{2M} \left ( \left (\nabla
\sqrt{n({\bf
r})}\right )^2 + (\nabla \zeta({\bf r}))^2 n({\bf r}) \right ) \nonumber \\
&-& \int d^3r \left [\left (\mu-U({\bf r}) \right )n({\bf r})-
\frac{n^2({\bf r})}{2} \left(c_0 + c_2 \langle {\bf F}({\bf r})
\rangle^2 \right ) \right ] .
\end{eqnarray*}
In this expression, $\mu$ is the chemical potential and the
averaged single-atom spin angular momentum is
\begin{equation}
\langle {\bf F}({\bf r})\rangle =
\sum_{\alpha,\beta}\zeta_\alpha^\star ({\bf r}){\bf
F}_{\alpha,\beta} \zeta_\beta({\bf r}).
\label{faverage}
\end{equation}

For $c_2 >0$, the energy $E$ is minimized by $\langle {\bf F}({\bf
r})\rangle =0$, and the spinor condensate is in an
``anti-ferromagnetic'', or ``polar'' state. This is the case for
$^{23}$Na condensates, in which case $a_2-a_0 \simeq 5$ a.u.
Ketterle and coworkers have studied this situation in great detail
\cite{mit}. In particular they have obtained spin-domain diagrams
and studied experimentally the miscibility of these domains in the
presence of external fields.

For $c_2 <0$, in contrast, the energy $E$ is minimized by making
$\langle {\bf F}({\bf r})\rangle^2 =1$. As discussed in Ref.
\cite{ho2}, the direction of the spin is
\begin{equation}
\langle {\bf F}({\bf r}) \rangle = \cos \beta_0 {\hat {\bf z}} +
\sin \beta_0 \times (\cos \alpha_0 {\hat {\bf x}} + \sin \alpha_0
{\hat {\bf y}}),
\end{equation}
where $\alpha_0$ and $\beta_0$ are Euler angles. All possible
orientations $(\alpha_0, \beta_0)$ are possible and lead to the
same ground-state energy $E$. Recent theoretical calculations by
Klausen {\em et al.} predict that for spin-1 $^{87}$Rb, the
scattering lengths $a_0$ and $a_2$ are almost equal, but with $a_0
> a_2$, with a difference of the order of 0.3 to 2.7 a.u. \cite{klausen}.

Consider, then, an $^{87}$Rb condensate trapped on an optical
lattice with wells deep enough that its ground state is the
Mott-insulator state, i.e., there is no global phase of the
condensate over many lattice sites \cite{zoller}. Each lattice
site is therefore the location of a ``mini-condensate,'' which can
contain as many as several thousands atoms in one-dimensional
lattices, and several hundreds in 2-D lattices. In the absence of
external fields and long-range site-to-site interactions, these
condensates can be thought of as independent magnets, whose spin
vectors point in random directions, with no spin correlations
between sites. This situation is similar to the spin lattices
familiar from the study of magnetism, with two differences. First,
the quantum mechanical exchange interaction, which is at the core
of magnetism, is completely negligible in the present situation.
This is because neighboring sites on an optical lattice are at
least one half optical wavelength apart. For deep lattice wells,
the center-of-mass wave functions for the individual
mini-condensate
--- essentially the ground state Wannier wave functions at the
individual sites --- do not have any significant overlap. Second,
the magnetic dipolar coupling, which is normally negligible and
leads to the prediction of Curie temperatures several orders of
magnitude lower than actually observed in solid state magnetic
materials, is now the dominant interaction, due to the large
number $N$ of atoms at each lattice site. This leads to an $N^2$
enhancement factor, as we now show.

\subsection{Magnetic dipole-dipole interaction}

In order to describe the magnetic dipolar interaction between
mini-condensates at lattice sites ${\bf i}$ and ${\bf j}$, we
assume that the condensates at each site can be treated
independently, and have the same spatial form, which is also
independent of the spin state of the atoms. Specifically, we
decompose the Schr{\"o}dinger field operator as
\begin{equation}
\psi({\bf r}) = \sum_{\alpha=0,\pm1} \psi_\alpha({\bf
r})|f=1,m_f=\alpha\rangle,\nonumber
\end{equation}
with
\begin{equation}
\psi_a({\bf r}) = \sum_{\bf i} \phi_i({\bf r}) {\hat
a}_\alpha({\bf i}).
\label{wf1}
\end{equation}
In this expansion, which goes beyond the mean-field approximation
of Eq. (\ref{wf}), ${\bf r}_{\bf i}$ is the coordinate of the
${\bf i}$-th lattice site, ${\hat a}_\alpha({\bf i})$ and ${\hat
a}^\dagger_\alpha({\bf i})$ are bosonic annihilation and creation
operators for atoms in the hyperfine state $\alpha$ at site $\bf
i$, and $\phi_i({\bf r})=\phi({\bf r-\bf r_i})$ is the ground
state wave function of the mini-condensate at that site,
normalized to unity. For $a_0 \simeq a_2$, it is approximately
given by the solution of the stationary Gross-Pitaevskii equation
\begin{equation}
\left [ -\frac{\hbar^2 \nabla^2}{2M} + U_{\bf i}({\bf r}) +
c_0(N_{\bf i}-1)|\phi_i({\bf r})|^2 -\mu \right ] \phi_i({\bf
r})=0, \nonumber
\end{equation}
where
\begin{equation}
N_{\bf i}=\sum_\alpha \langle{\hat a}_\alpha^\dagger({\bf i}){\hat
a}_\alpha({\bf i})\rangle , \nonumber
\end{equation}
is the total number of atoms at site ${\bf i}$ and we assume that
all sites have the same number of atoms.

The magnetic dipole-dipole interaction between the
mini-condensates at sites ${\bf i}$ and ${\bf j}$ is given by
\cite{add}
\begin{eqnarray}
V_{dd}^{{\bf i}{\bf j}}  = \frac{\mu_0}{4\pi} \int d^3r \int
d^3r'\, |\phi ({\bf r}-{\bf r_i})|^2|\phi ({\bf r'}-{\bf r_j})|^2
\nonumber \\
\times \left[\frac{\vec{\mu}_{\bf i} \cdot \vec{\mu}_{\bf
j}}{|{\bf {r}}-{\bf r}'|^3}-\frac{3(\vec{\mu}_{\bf i} \cdot ({\bf
r}-{\bf r}') ) ( \vec{\mu}_{\bf j} \cdot ({\bf r}-{\bf
r}'))}{|{\bf {r}}-{\bf r}'|^5} \right] , \nonumber
\end{eqnarray}
where $\mu_0$ is the vacuum permeability and $\vec{\mu}_{{\bf i}}$
is the magnetic dipole moment at site ${\bf i}$. In
second-quantized form, it is given explicitly by
\begin{equation}
\vec{\mu}_{{\bf i}}= \gamma_B \sum_{\alpha, \beta}
\hat{a}^{\dagger}_{\alpha}({\bf i}) {\bf F}_{\alpha, \beta}
\hat{a}_{\beta}({\bf i}) \equiv \gamma_B {\bf S}_{{\bf i}},
\nonumber
\end{equation}
where $\gamma_B=g_F\mu_B$ is the gyromagnetic ratio and we
recognize that ${\bf S}_{{\bf i}}$ is the angular momentum
operator for the condensate at site ${\bf i}$. We remark that for
a given site, the expectation value of $\vec{\mu}_{{\bf i}}$ is
\begin{eqnarray}
\langle {\vec \mu}_{\bf i} \rangle &=& \gamma_B \sum_{\alpha,
\beta} \langle {\hat a}^\dagger_\alpha({\bf i}) {\bf F}_{\alpha,
\beta}
{\hat a}_\beta({\bf i}) \rangle  \nonumber \\
&\simeq& N_{\bf i} \gamma_B \langle {\bf F}_{\bf i} \rangle,
\nonumber
\end{eqnarray}
where $\langle {\bf F}_{\bf i} \rangle$ is the single-atom
magnetization at the site, see Eq. (\ref{faverage}).

Summarizing, then, the Hamiltonian describing the spinor
``mini-condensates" in the optical lattice, subject to
spin-changing collisions and to an inter-site magnetic dipolar
interaction has the spin-spin coupling form
\begin{eqnarray}
H=&&\sum_{\bf i} \left[ \lambda_a' {\bf S}_{\bf i}^2  +  \gamma_B
\sum_{{\bf j} \neq {\bf i}} \lambda_{{\bf i}{\bf j}}
{\bf S}_{\bf i} \cdot {\bf S}_{\bf j} \right. \nonumber \\
&& \left. - 3\gamma_B \sum_{{\bf j} \neq {\bf i}}{\bf S}_{\bf i}
\cdot  {\bf {\Lambda}}_{{\bf i}{\bf j}}  \cdot {\bf S}_{\bf j} -
\gamma_B {\bf S}_{\bf i} \cdot {\bf B}_{\rm ext}\right], \label{h}
\end{eqnarray}
where
\begin{eqnarray}
\lambda_a' &=& (1/2)c_2 \int d^3r|\phi ({\bf r}-{\bf r_i})|^4, \nonumber \\
\lambda_{{\bf i}{\bf j}} &=& \frac{\gamma_B \mu_0}{4 \pi} \int
d^3r \int d^3r'\frac{|\phi ({\bf r}-{\bf r_i})|^2|\phi ({\bf
r'}-{\bf r_j})|^2 }{|{\bf {r}}-{\bf r}'|^3} \nonumber
\end{eqnarray}
and the tensor ${\bf {\Lambda}}_{{\bf i}{\bf j}}$ is defined by
\[
{\bf{\Lambda}}_{{\bf i}{\bf j}}=\frac{\gamma_B \mu_0}{4 \pi} \int
d^3r \int d^3r'\frac{|\phi ({\bf r}-{\bf r_i})|^2|\phi ({\bf
r'}-{\bf r_j})|^2 ({\bf r}-{\bf r}')^2}{|{\bf {r}}-{\bf r}'|^5}.
\]
We have also introduced an external magnetic field ${\bf B}_{\rm
ext}$ for future use. In the limit of tight confinement, the
condensate wave functions at each lattice site can be approximated
by
$$
|\phi ({\bf r}-{\bf r_i})|^2\approx \delta({\bf r}-{\bf r_i}).
$$
In this limit we have
$$
\lambda_{{\bf i}{\bf j}}=\frac{\gamma_B \mu_0} {4 \pi |{\bf
r}_{{\bf i}{\bf j}}|^3},
$$
and the tensor $\bf{\Lambda}_{{\bf i}{\bf j}}$ becomes
$$
{\bf {\Lambda}}_{{\bf i}{\bf j}}=\lambda_{{\bf i}{\bf j}}{\bf
{\hat{r}}}_{{\bf i}{\bf j}}^2,
$$
where ${\bf {r}}_{{\bf i}{\bf j}} = {\bf r}_{\bf i} - {\bf r}_{\bf
j}$ and ${\bf {\hat{r}}}_{{\bf i}{\bf j}}= {\bf {r}}_{{\bf i}{\bf
j}}/ |{\bf {r}}_{{\bf i}{\bf j}}|$.

\section{Ferromagnetism in a 1D optical lattice}

In this section, we study the magnetic properties and spin
dynamics of spinor condensates in a 1D optical lattice. More
specifically, we consider a blue-detuned optical lattice where the
mini-condensates are trapped at the standing wave nodes. In this
case, the light-induced dipolar interaction can be ignored and the
mini-condensates only interact via the magnetic dipolar
interaction. Without loss of generality, we assume that the axis
of the lattice is along the z direction, which we also choose as
the quantization axis. Hence the total Hamiltonian (\ref{h})
reduces to
\begin{eqnarray}
H=&&\sum_i \left[ \lambda_a' {\bf S}_i^2  +  \gamma_B \sum_{j \neq
i} \lambda_{ij}
{\bf S}_i \cdot {\bf S}_j \right. \nonumber \\
&& \left. - 3\gamma_B \sum_{j \neq i} \lambda_{ij} S_i^z S_j^z  -
\gamma_B {\bf S}_i \cdot {\bf B}_{\rm ext}\right]. \label{h1}
\end{eqnarray}

We assume that the magnetic field ${\bf B}_{\rm ext}$ is of the
form
\[ {\bf B}_{\rm ext}=B_z \hat{{\bf z}} + B_{x} \hat{{\bf x}}, \]
where $B_z \hat{{\bf z}}$ is an applied field and $B_{x} \hat{{\bf
x}}$ is an effective magnetic field that accounts for all possible
effects from the experimental environment. While this field can
have any possible orientation, we take it to be transverse and
along ${\bf \hat{x}}$ without loss of generality, since any
longitudinal component can be included in $B_z$.

Furthermore, we consider an infinitely long lattice so that
boundary effects can be ignored. The hamiltonian describing the
spin ${\bf S}$ of a generic site $i$ reads then
\begin{eqnarray}
h=&& \lambda_a' {\bf S}^2 - \gamma_B {\bf S} \cdot \left[
\left(B_z+2 \sum_{j \neq
i} \lambda_{ij} S_j^z \right) \hat{\bf z} \right. \nonumber \\
&& + \left. \left(B_{x} - \sum_{j \neq i} \lambda_{ij} S_j^{x}
\right) \hat{\bf x}  - \sum_{j \neq i} \lambda_{ij} S_j^{y}
\hat{\bf y}\right] . \label{hi}
\end{eqnarray}
We now proceed to determine the ground state of the single-site
Hamiltonian (\ref{hi}) in the mean-field --- or Weiss molecular
field --- approximation \cite{solid}. It consists in replacing the
operators $S_j^{\alpha}$, $\alpha =x, y, z$, by their ground-state
expectation value
\begin{equation}
\langle S_j^{\alpha} \rangle \rightarrow M_{\alpha} =N m_{\alpha}
,
\end{equation}
which is assumed to be the same for all sites. We remark that
$m_z$ is nothing but the difference in population of the Zeeman
sublevels of magnetic quantum numbers $\pm 1$. Replacing
$S_j^{\alpha}$ by $N m_{\alpha}$ allows us to approximate the
Hamiltonian (\ref{hi}) by
\begin{equation}
h_{\rm {mf}} =  \lambda_a' {\bf S}^2- \gamma_B {\bf S} \cdot {\bf
B}_{{\rm eff}}, \label{heff}
\end{equation}
where we have introduced the effective magnetic field
\begin{equation}
{\bf B}_{\rm {eff}} = (B_z+2\Lambda m_z) \hat{\bf z} + (B_x -
\Lambda m_{x}) \hat{\bf x}-\Lambda m_{y} \hat{\bf y}, \nonumber
\end{equation}
and \[ \Lambda = N \sum_{j\neq i} \lambda_{ij} . \]

In the case of $^{87}$Rb, the individual spinor condensates at the
lattice sites are ferromagnetic, $\lambda_a' <0$.  In that case,
the ground state of the mean-field Hamiltonian (\ref{heff}) must
correspond to a situation where the condensate at the site $i$
under consideration must be aligned along ${\bf B}_{{\rm eff}}$
and takes its maximum possible value $N$. That is, the ground
state of the mean-field Hamiltonian (\ref{heff}) is simply
\begin{equation}
|GS \rangle = |N, N \rangle_{{\bf B}_{{\rm eff}}}, \label{gs}
\end{equation}
where the first number denotes the total angular momentum and the
second its component along the direction of ${\bf B}_{{\rm eff}}$.
Note that $|GS \rangle$ represents a spin coherent state in the
basis of $|S,S_z \rangle$. The fact that the ground state magnetic
dipole moment of each lattice site is $N$ times that of an
individual atom results in a significant magnetic dipole-dipole
interaction even for lattice points separated by hundreds of
nanometers. This feature, which can be interpreted as a signature
of Bose enhancement, is in stark contrast with usual
ferromagnetism, where the magnetic interaction is negligible
compared to exchange and where the use of fermions is essential.

The mean-field ground state of Eq.~(\ref{gs}) allows us to
calculate the magnetization $m_{x,y,z}$. One finds readily
\[ m_\alpha = \frac{1}{N} \langle GS | S_i^\alpha |
GS  \rangle =  \cos \theta_\alpha ,\] where $\theta_\alpha$ is the
angle between ${\bf B}_{{\rm eff}}$ and the $\alpha$-axis. In the
absence of externally applied field, $B_z = 0$, this gives
\begin{subequations}
\label{mxz}
\begin{eqnarray}
m_z &=& \frac{ 2 \Lambda m_z}{B},\\
 m_{x} &=& \frac{B_{x}- \Lambda
m_{x}}{B},\\
m_{y} &=& -\frac{\Lambda m_y}{B}
\end{eqnarray}
\end{subequations}
where $B=\sqrt{( 2 \Lambda m_z)^2+(B_{x}- \Lambda
m_{x})^2+(\Lambda m_y)^2}$ normalizes the magnetization vector to
unity .

Since $B>0$, the third of these equations implies that $m_y=0$.
With $m_x^2+m_z^2=1$ and the condition $2\Lambda=B$, which follows
directly from the equation for $m_z$, we find further that for
$B_{x} \ge 3\Lambda$, the unique solution is $m_z=m_y=0$, $m_x=1$.
That is, the lattice of condensates is magnetically polarized
along the environmental magnetic field $B_x$. For $B_x <
3\Lambda$, in contrast, there are two coexisting sets of
solutions: i) $m_z=m_y=0$ and $m_{x}=1$; and ii) $m_z=\pm
\sqrt{1-(B_{x}/ 3\Lambda)^2}$, $m_y=0$ and $m_{x}=B_{x}/3\Lambda$.
It is easily seen that the state associated with the latter
solutions has the lower energy . Hence it corresponds to the true
ground state, while solution 1 represents an unstable equilibrium.

We have, then, the following situation: As the effective magnetic
field strength $B_x$ is reduced below the critical value $3
\Lambda$, the lattice ceases to be polarized along the direction
of that field. A phase transition occurs, and a {\em spontaneous
magnetization} along the $z$-direction appears, characterized by a
finite $m_z$. This phenomenon is reminiscent of conventional
ferromagnetism. Indeed, our model is analogous to the Ising
model\cite{huang}, with the environmental transverse magnetic
field $B_{x}$ playing the role of temperature. For $B_{x}=0$ ---
corresponding to zero temperature in Ising model --- the spins at
each lattice site ${\bf S}_i$ align themselves along the lattice
direction, even in the absence of longitudinal field. This
spontaneous spin magnetization diminishes as $B_{x}$ increases,
and completely vanishes if $B_{x}$ exceeds the critical value
$3\Lambda$ --- the analog of the Curie temperature in the Ising
model. We note however that the situation at hand exhibits
important qualitative differences with the Ising model. For
example, no spontaneous magnetization occurs in 1D Ising model,
for any finite temperature.

We note however that the appearance of a spontaneous magnetization
does not rely on this condition being fulfilled. This point was
discussed in Ref.\cite{han}, which numerically solved the
Hamiltonian (\ref{h1}) without invoking the mean-field
approximation for a two-well system and showed how the situation
rapidly approaches the mean-field results as $N$ increases.

\section{Anti-ferromagnetic Ground State of the 2D Lattice}
Now we turn our attention to 2D lattices, formed as before by
blue-detuned lasers. We show that depending on the relative
magnitude of the lattice constants along its two axes, this system
exhibits a variety of possible ground states, including an
anti-ferromagnetic configuration.

We consider a rectangular lattice in the $(y,z)$-plane, with
primitive lattice vectors ${\bf a}=a \hat{\bf z}$ and ${\bf b}=b
\hat{\bf y} $, of lengths $a$ and $b$, in these two directions. We
assume as before that the number of atoms at each lattice site is
the same and that the atoms are tightly confined so that we can
approximate their probability density by a delta function at each
lattice site,
$$
|\phi_{ij}({\bf r})|^2 =\delta({\bf r}-{\bf r}_{ij}).
$$
Here, ${\bf r}_{ij}=i {\bf a} + j {\bf b}$ is the position of the
center of the $(i,j)$ lattice site. Under these conditions, the
Hamiltonian~(\ref{h}) with ${\bf B}_{\rm ext}=0$ becomes
\begin{eqnarray}
H=&&\sum_{ij} \left[ \frac{\lambda_a}{2} {\bf S}_{ij}^2  +
 \frac{\gamma_B \mu_0}{4\pi} \sum_{kl \neq ij}
{\bf S}_{ij}^{T} \cdot \Lambda_{ij,kl} \cdot {\bf S}_{kl} \right]
\label{h11}
\end{eqnarray}
where
\begin{equation}
\Lambda_{ij,kl} =  \left(
\begin{array}{ccc} \frac{1}{|{\bf r}_{ij,kl}|^3} & 0 & 0
\\ 0 & \frac{1}{|{\bf r}_{ij,kl}|^3} -\frac{3(na)^2}{|{\bf r}_{ij,kl}|^5}
& -\frac{3(na)(mb)}{|{\bf r}_{ij,kl}|^5}
\\ 0 & -\frac{3(na)(mb)}{|{\bf r}_{ij,kl}|^5}
& \frac{1}{|{\bf r}_{ij,kl}|^3} -\frac{3(mb)^2}{|{\bf
r}_{ij,kl}|^5}
\end{array} \right).  \label{Matrix}
\end{equation}
and  ${\bf r}_{ij,kl}={\bf r}_{ij}-{\bf r}_{kl}  = n {\bf a} + m
{\bf b}$, with $n = i-k$ and $m = j-l$.

\subsection{infinite size lattices}

As in the preceding section, we determine the ground state of the
lattice in the semiclassical limit, ignoring spin-spin
correlations and replacing the operators ${\bf S}_{ij}$ with their
expectation value with respect to a spin coherent state,
\[
{\bf S}_{ij}\rightarrow\langle{\bf S}_{ij}\rangle={\bf M}_{ij}.
\]
The semiclassical ground state corresponds to the orientation of
the spin vectors that minimizes the semiclassical energy
corresponding to the Hamiltonian (\ref{h11}). In contrast to the
one-dimensional case, it is not obvious from inspection of
Eq.~(\ref{h11}) that all expectation values  ${\bf M}_{ij}$ should
be equal. Hence, the determination of the ground state for an $N
\times M$ lattice requires the minimalization of the energy with
respect to $2NM$ variables. However, in the limit of an infinite
lattice the ground state should be translationally invariant with
respect to displacements of the spins by a finite number of
lattice constants along either axis. We can therefore generalize
the mean-field ansatz used in the one-dimensional case by assuming
that the 2D lattice can be decomposed into a finite set $\{ \ell
\}$ of interpenetrating periodic sublattices for which all spin
vectors have the same orientation.

The positions of the sites of the sublattice $\ell$ of primitive
lattice vectors ${\bf a}_\ell$ and ${\bf b}_\ell$ are
\[
{\bf r}_{\ell,ij}=i {\bf a}_\ell + {\bf a}_{\ell,0}+ j {\bf
b}_\ell + {\bf b}_{\ell,0}
\]
where ${\bf a}_{\ell,0}$ and ${\bf b}_{\ell,0}$ denote the origin
of that lattice. Since the interaction between dipole moments that
are perpendicular to the plane of the lattice is repulsive while
the interaction between dipole moments in the plane of the lattice
is predominantly attractive, the ground state must correspond to
spin vectors in the plane $(y,z)$ of the lattice. Hence the spin
vector associated with the sublattice $\ell$ can be written as
\[
{\bf M}_{\ell}=N\left (\cos\theta_\ell\hat{{\bf y}}+
\sin\theta_\ell\hat{{\bf z}} \right).
\]

One can gain an intuitive feel for the ground state of the system
by considering what happens when one lets a 1D lattice approach an
already existing one from infinity. For concreteness, we take the
axes of both lattices to be along $\hat{{\bf z}}$. We know from
the previous section that for large lattice separations, the spins
in each lattice will be oriented in either the $+\hat{{\bf z}}$ or
$-\hat{{\bf z}}$ direction with equal probability. In effect, each
lattice acts like a long bar magnet. As the lattices approach each
other, though, they start to interact via their magnetic dipole
moments. The minimization of energy then proceeds in a familiar
way:  Just as two bar magnets placed side by side orient
themselves so that opposite poles are next to each other, the
spins of the two 1D lattices will arrange their orientation so
that the spins in one lattice point along $+\hat{{\bf z}}$ while
in the other lattice the spins point along $-\hat{{\bf z}}$. This
will remain true as long as the lattice separation is much larger
than the primitive lattice vector ${\bf b}$ of the 1D lattices,
due to the $1/r^3$ dependence of the magnetic dipole interaction.
Indeed, in this case the easy axis is the $y$-axis.

This argument can easily be generalized to many rows. It follows
that for $a \gg b$, rows of spins parallel to the $z$-axis will
alternatively align themselves along the $+\hat{{\bf z}}$ and
$-\hat{{\bf z}}$ direction. Similarly, for $a \ll b$, the $z$-axis
becomes the easy axis and rows of spins parallel to $y$ align
themselves alternatively along the $+\hat{{\bf y}}$ or $-\hat{{\bf
y}}$ direction. In both cases, though, the ground state is
expected to be anti-ferromagnetic.

\begin{figure}
\includegraphics*[width=.72\columnwidth,height=.72\columnwidth]{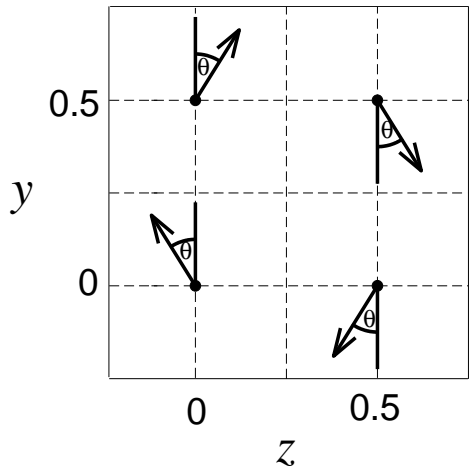}
\caption{Orientations of the spins on the four interpenetrating
sublattices for $a=b$. The lengths are in units of
$\lambda$.}\label{4sub}
\end{figure}

Even though the magnetic dipole interaction is long ranged, it is
easy to see that neighboring spins within each row interact more
strongly than do neighboring spins in adjacent rows provided that
$a=b+\epsilon$ with $\epsilon$ positive. One therefore expects
that the ground state will remain anti-ferromagnetic unless
$\epsilon\rightarrow 0$. In this limit, there are clearly two
degenerate anti-ferromagnetic ground states that are topologically
distinct, i.e. that can not be related by a simple rotation. Any
weighted combination of these two configurations has the same
energy and is therefore a new degenerate ground state. Assigning
the weight $\cos^2\theta$ to the ground state with all spins
pointing in the $\pm{\bf \hat{y}}$ direction and $\sin^2\theta$ to
the ground state with all spins in the $\pm{\bf \hat{z}}$, then we
find that this situation is equivalent to a ground state
consisting of four interpenetrating sublattices ($\ell=1,2,3,4$),
with spin orientations,
\begin{subequations}
\label{M1}
\begin{eqnarray}
M_1 &=& N(\cos\theta {\bf \hat{y}}+\sin\theta{\bf \hat{z}}) , \\
M_2 &=& N(-\cos\theta {\bf \hat{y}}+\sin\theta{\bf \hat{z}}) ,\\
M_3 &=& N(-\cos\theta {\bf \hat{y}}-\sin\theta{\bf \hat{z}}) ,\\
M_4 &=& N(\cos\theta {\bf \hat{y}}-\sin\theta{\bf \hat{z}}),
\end{eqnarray}
\end{subequations}
and sublattice sites located at
\begin{subequations}
\label{r4}
\begin{eqnarray}
{\bf r}_{1,ij} &=& 2i {\bf a}+2j {\bf b} ,\\
{\bf r}_{2,ij} &=& 2i {\bf a}+(2j+1) {\bf b} ,\\
{\bf r}_{3,ij} &=& (2i+1){\bf a}+(2j+1) {\bf b} ,\\
{\bf r}_{4,ij} &=& (2i+1){\bf a}+2j {\bf b} ,
\end{eqnarray}
\end{subequations}
where $i,j=0,\pm 1,\pm 2,...$. The corresponding lattice structure
is illustrated in Fig. \ref{4sub}. For $a= b$ all values of
$\theta$ are degenerate while for $a < b$ and $a
> b$ the ground state corresponds to $\theta$ equal to $0$
and $\theta=\pi/2$, respectively.

The next section discusses the use of a genetic algorithm to
numerically determine of the lattice ground state for a finite
lattice size.

\subsection{Finite size lattices}

\subsubsection{Genetic algorithm}

Genetic algorithms have become a widely used tool for solving
optimization problems that depend on a large number of variables
\cite{geneticalgorithms}. The basic idea behind genetic algorithms
is Darwinian natural selection. These algorithms proceed from an
initial set of trial solutions to the optimization problem, which
can be thought of as individuals in a population.  The individuals
breed, following some prescribed mating rules, to produce
offspring, which constitute the next generation of individuals. In
addition, random mutations are also introduced. The offspring that
produce better solutions to the problem survive and are allowed to
further breed, while those that produce poor solutions are
eliminated. Ideally, after many generations the algorithm
converges to the optimal solution(s) to the problem at hand.

In the specific system at hand, the algorithm starts from a large
population ${\cal N}$ of initial lattices, typically ${\cal
N}=512$. Most of them have completely random spin orientations,
but some may have ordered configurations based on the ground state
of the infinite lattice. At each generation, the genetic algorithm
performs a combination of mutations and breeding steps on the
members of the population, which we refer to as mutating and
mating.

The mutations modify each member of the population to form a
second population of ${\cal N}$ lattices. They can be either
global and local. Local mutations involve giving random rotations
to a random percentage of the spins in the individual lattices.
These rotations are by angles $\varphi$ and $\theta$ about the $y$
and $x$ axes, respectively, where $\varphi$ and $\theta$ are
normally distributed random numbers with standard deviations
typically chosen to be $\pi/8$. In contrast, the global mutations
rotate {\it all} spin in the lattice by related amounts: They
either apply the same random rotation to all lattice sites, or
rotate the spin at each lattice site by a slightly different
amount determined by its value (this is used when investigating
the case of equal lattice constants, $a = b$, discussed below). In
general, a given individual is subjected to both local and global
mutations.

After the mutations are performed, the $2{\cal N}$ individuals are
allowed to mate. The mating process randomly picks two individuals
using a normally distributed probability distribution centered
around individuals with the lowest energy. This insures that, on
average, only those individuals with the lowest energies produce
offspring. Each pair of parents produces two offsprings using one
of four randomly selected mating techniques: site swapping,
sub-lattice swapping, row and column swapping, and row and column
rearranging. Site swapping consists of swapping a random number of
randomly chosen sites from the parents. Similarly, sub-lattice
swapping consists of swapping a randomly sized and positioned
sub-lattice between the parents. Row and column swapping works by
randomly picking rows from both parents and forming one child, and
doing the same with columns to form a second child. Row and column
rearranging uses only a single parent to produce a child by
randomly rearranging its rows or columns. The mating process is
repeated ${\cal N}$ times at each generation to produce a total
population of $4{\cal N}$ lattices. Of those, only the ${\cal N}$
individuals with the lowest energy are selected as parents for the
next generation.

The genetic algorithm is run until the relative energies of the
individuals in generation ${\cal M}$ and ${\cal M}-100$ differ by
less than $10^{-7}$.

\subsubsection{Numerical results}

The ground state of the system determined by the genetic algorithm
is characterized by all spins lying in the plane of the lattice,
in agreement with the discussion of section IV. If the lengths of
the primitive lattice vectors $a$ and $b$ differ significantly,
say, by 10 percent or more, the ground state is
anti-ferromagnetic. With the exception of sites near the lattice
boundary, the anti-ferromagnetic structure is identical to that
predicted based on an infinite lattice.

As is to be expected, boundary effects become more important, the
smaller the lattice. In that case, the ground state is
characterized by spins orientations near the boundaries that
deviate from the $\pm {\bf \hat{y}}$ or $\pm {\bf \hat{z}}$
directions. When $a$ and $b$ are significantly different, these
boundary effects are manifest only near the corners of the
lattice, and they lower the ground state energy by a very small
amount. For example, for $a = 0.6\lambda$ and $b = 0.5\lambda$,
where $\lambda$ is the wavelength of the laser forming the lattice
in the $z$-direction, the boundary effects reduce the ground-state
energy of an $11\times 11$ lattice by only $0.1$ percent compared
to its infinite lattice value. For larger lattices, the boundary
effects become even smaller.

\begin{figure}
\includegraphics*[width=.72\columnwidth,height=1.5\columnwidth]{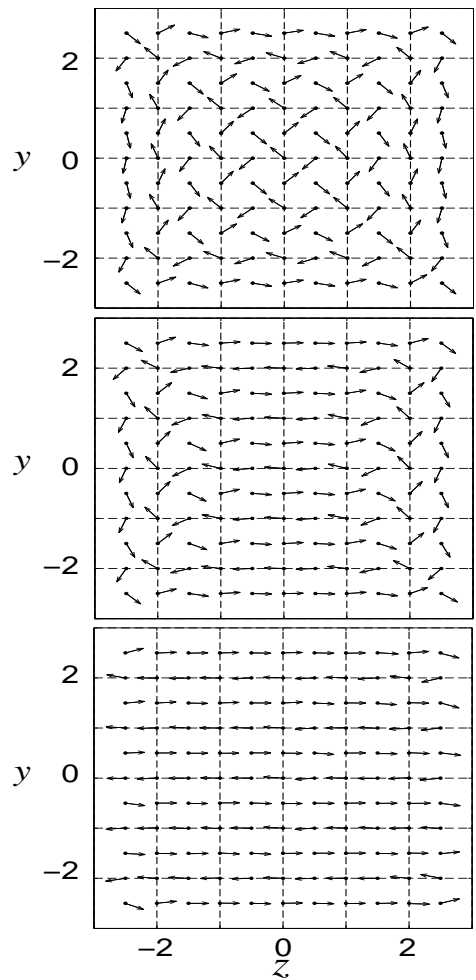}
\caption{The ground state configuration shows the transition from
the boundary dominated pattern for $a=b$ to the anti-ferromagnetic
configuration for $a \neq b$. From top to bottom, $b=0.5$, 0.505
and 0.6, respectively and $a=0.5$ for all figures. The lengths are
in units of $\lambda$. }\label{fig2}
\end{figure}

When $a=b$, finite size effects are more important in determining
the spin structure of the ground state. We recall that in that
case, an infinite lattice possesses an infinite number of
degenerate ground states characterized by the angle $\theta$.
Boundary effects break this degeneracy and lead to the appearance
of a preferred pattern. Fig.~\ref{fig2} illustrates the transition
from the boundary dominated pattern of the $a=b$ situation to the
anti-ferromagnetic configuration of $a \neq b$. As illustrated in
Fig.~\ref{fig2}a, for the case of $a=b$, near boundaries the spins
are aligned parallel to them. That this should be the case is
plausible since when going from a situation where $a<b$ to $a>b$,
the spin orientation must go from being parallel to the $y$-axis
to being parallel to the $z$-axis. To accommodate the orthogonal
directions along two adjacent boundaries, the angle $\theta$ near
the corners changes in such a way that the spins at the corner
sites  make an angle of $\pi/4$ relative to the $y$- and $z$-axis.
This lifts the degeneracy present in the infinite lattice. As a
result, the spins near the center of the finite lattice always
take on an orientation corresponding to Eqs.~(\ref{M1}) and
(\ref{r4}) with $\theta=\pi/4$. This result holds for all
finite-size lattices. Finally, we note that the ground-states of
finite-size lattices are two-fold degenerate, the second ground
state being obtained by reflections about the $y$ and $z$ axes.

\begin{figure}
\includegraphics*[width=.65\columnwidth,height=.6\columnwidth]{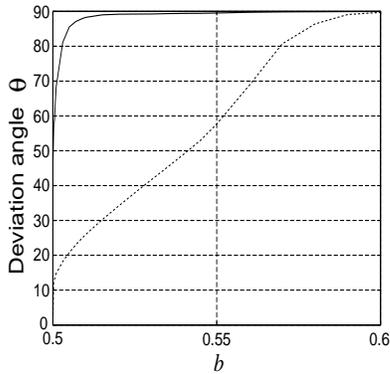}
\caption{Plot of the deviation angle $\theta$ relative to the $y$
axis for the spin at the center of the lattice (solid curve) and a
spin on left boundary of the lattice (dashed curve) as functions
of $b$ for fixed $a=0.5$. Lengths are in units of $\lambda$.
}\label{fig3}
\end{figure}

Figure \ref{fig3} shows how the spins orients themselves as $b$
changes for fixed $a$. As $b$ deviates from $a$, the spins near
the center of the lattice quickly become parallel to the
easy-axis, while the spins near the boundaries become so much more
slowly.

\section{summary and outlook}
In summary, we have studied the spin configurations and magnetic
properties of spinor Bose-Einstein condensates in an optical
lattice. In the tight-binding limit, the ground state is the
Mott-insulator state and the condensed atoms at each lattice site
collectively behave as a spin magnet. Due to Bose enhancement, the
dipole-dipole interactions between these spin magnets become
important and may give rise to a rich variety of phenomena. We
have shown here that the array of spin magnets can undergo a
ferromagnetic (in the 1D case) or anti-ferromagnetic (in the 2D
case) phase transition under the dipolar interaction when external
magnetic fields are sufficiently weak. Using the same mechanism,
it will also be possible to create ferrimagnetic lattice systems
if one can interleave two sets of optical dipole potentials, each
trapping one species of atoms (or one hyperfine state of the same
atom) different from the other.

In the case of a far red-detuned lattice such that the spacing
between adjacent lattice site exceeds the atomic resonant
wavelength, the detection of the ground state spin structure
amounts to detecting populations in the individual Zeeman
sublevels at each site. This can be achieved using a Raman
scattering scheme. For example, one can shine two light beams, one
$\pi$-polarized and the other circularly polarized, onto the
system. The absorption or gain of the probing light after passing
the sample is then a measure of the relative population of the
hyperfine levels, since it depends upon which of them are
initially populated. This scheme wouldn't work for a blue detuned
lattice, though, since in that case the spacing between
neighboring sites is sub-wavelength. However, the long range
periodic spin structure, in particular the ferromagnetic and
anti-ferromagnetic ordering, can still be detected by Bragg
scattering \cite{bragg}. Let us take $^{87}$Rb as an example. Its
ground state is the 5S$_{1/2}$ state with $F=1$. For
$\sigma^+$-polarized Bragg probe light (we choose the quantization
axis to be parallel or antiparallel to atomic spins) with a
frequency close to the $F=1 \rightarrow F'=2$ D2 resonance line ,
then the ratio of the transition strength (or scattering cross
section) for atoms in $m=-1$ and $m=1$ Zeeman sublevel is $1/6$.
As a result, the Bragg signal depends on whether one has a
ferromagnetic lattice (where all the atoms are in either $m=-1$ or
$m=1$ Zeeman sublevel) or anti-ferromagnetic lattice (where half
the atoms are in $m=-1$ and the other half are in $m=1$ sublevel).

In addition to their ground state structure, spinor condensates in
an optical lattices also possesses considerable potential for
studying other phenomena such as spin waves \cite{spinwaves},
macroscopic magnetization tunneling \cite{tunneling}, domain wall
formation, etc. Future studies will also include the dynamical
properties of the system. Due to the long-range as well as the
nonlinear nature of the dipolar interaction, the dynamics of the
system should be very rich. For instance, given a ground state 2D
lattice with primitive lattice constants $a<b$ where all the spins
are aligned along the $\pm \hat{\bf y}$ direction, one can
suddenly modify the lattice light so that $a>b$. Whether and how
the spins adjust themselves to the new ground state will be an
interesting problem, closely related to the phenomenon of spin
tunneling \cite{tunneling}. In addition, these systems may also
find applications in the field of quantum information and
computation. We conclude by noting that in addition to the Mott
insulator limit studied in this paper, the genetic algorithm that
we have developed here might be modified to investigate the other
limit where tunneling between lattice sites becomes significant
and the system becomes a superfluid \cite{goral}.

\acknowledgments We thank Prof. Poul Jessen for helpful discussion
and S. P\"{o}tting for help with the figures. This work is
supported in part by the US Office of Naval Research under
Contract No. 14-91-J1205, by the National Science Foundation under
Grants No. PHY00-98129, by the US Army Research Office, by NASA
Grant No. NAG8-1775, and by the Joint Services Optics Program.


\begin{references}
\bibitem{note} For a red-detuned lattice, the atoms are
trapped in the antinodes of lattice laser beams and for
appropriate laser parameters, the light-induced dipole-dipole
interaction can have an important effect on spin dynamics as
studied in Ref.\cite{spinwaves}.
\end{references}
\end{document}